\documentstyle[epsfig]{elsart}

\begin{document}
\hspace*{\fill}OCU-PHY-172, 1999
\runauthor{Dirks, Ni\'{e}gawa, Okano}
\begin{frontmatter}

\title{A Slavnov-Taylor identity 
and equality of damping rates for static transverse and longitudinal 
gluons in hot QCD} 
\author[osaka,bielefeld]{M. Dirks,\thanksref{Bielefeld}} 
\author[osaka]{A. Ni\'{e}gawa,\thanksref{email}} 
\author[osaka]{K. Okano\thanksref{okano}} 
\address[osaka]{Department of Physics, Osaka City University, 
Sumiyoshi-ku, Osaka 558-8585, Japan}
\address[bielefeld]{Fakult\"{a}t f\"{u}r Physik, Universit\"{a}t 
Bielefeld, D-33501 Bielefeld, Germany}
\thanks[Bielefeld]{Supported by the German Academic Exchange Service 
(DAAD). \\
Electronic address: dirks@sci.osaka-cu.ac.jp} 
\thanks[email]{Electronic address: niegawa@sci.osaka-cu.ac.jp} 
\thanks[okano]{Electronic address: okano@sci.osaka-cu.ac.jp} 
\date{Received today} 

\maketitle 
\begin{abstract} 
A Slavnov-Taylor identity is derived for the gluon polarization 
tensor in hot QCD. We evaluate its implications for damping of 
gluonic modes in the plasma. Applying the identity to next to the 
leading order in hard-thermal-loop resummed perturbation theory, we 
derive the expected equality of damping rates for static transverse 
and longitudinal (soft) gluons. This is of interest also in view of 
deviating recent reports of $\gamma_t(p=0)\neq\gamma_l(p=0)$ based 
on a direct calculation of $\gamma_l(p=0)$. 

\hspace*{1ex} 
\end{abstract} 
\end{frontmatter} 
\newpage 
\setcounter{equation}{0} 
\setcounter{section}{0} 
\section{Introduction} 
\def\theequation{\mbox{\arabic{equation}}} 
Much interest is devoted to the physics of a the deconfinement phase 
of hadronic matter (quark-gluon plasma, QGP), with both strong 
experimental and theoretical research going on. A good theoretical 
understanding of this new phase of matter is urgently required, in 
particular, in order to identify worthwhile observables of this new 
phase of matter. Insight can be gained in the framework of \lq Hot 
QCD' supplemented with a perturbative approach and important 
progress has been made during the last decade \cite{tft}. In 
particular thermal effects are known to alter soft modes in an 
important way, changing dispersion relations, and allowing for 
Landau damping as well as for new modes of propagation for both 
fermionic (quarks) and bosonic (gluons) quasiparticles \cite{rus}. 
The collective dynamics involved can be consistently described 
within Hard-Thermal-Loop (HTL) resummed perturbation theory 
\cite{htl,lb}. 

In this letter we are interested in damping of collective gluonic 
excitations as one of the most important characteristics of the 
plasma dynamics. While on-shell damping is known to be absent to HTL 
order $g^2 T^2$ a non zero damping rate is expected to arise at 
order $g(g^2T^2)$. In early work by Braaten and Pisarski \cite{bp} 
a finite, gauge independent result for the damping rate of static 
transverse gluons $\gamma_t(\omega = m_g, p = 0)$ has been obtained, 
as one of the important applications of the HTL-resummation scheme. 
Note that gauge invariance also follows from a general theorem 
\cite{kkr}. Also, in the static regime, $\gamma_l(m_g, p = 0) = 
\gamma_t(m_g, p = 0)$ has been mentioned in \cite{bp}, as is 
expected since longitudinal (l) and transverse (t) degrees of 
freedom can not be distinguished at zero spatial momentum. However, 
a complete understanding of the damping behavior in the plasma is 
not yet available. In particular, for non-zero values of the spatial 
momentum singular results are obtained for transverse and also for 
longitudinal gluons \cite{pis}, which originate from the absence of 
a chromo-magnetic mass in the HTL resummed gluon propagator. 
Moreover, also the analyticity of the damping rates around zero 
momentum has been questioned recently \cite{al}: Expansion for small 
$p$ of the imaginary part of the gluon polarization tensor has been 
reported to lead to $\gamma_t(m_g, 0) \neq \gamma_l(m_g, 0)$ and 
even to a quadratic divergence in $\gamma_l (m_g, p = 0)$. 

In this letter we derive a Slavnov-Taylor (ST) identity for the 
gluon polarization tensor in Coulomb gauge. After discussing some 
general consequences of it, we apply it to the next to leading order 
contribution to the gluon polarization tensor $\displaystyle{ 
\raisebox{0.6ex}{\scriptsize{*}}} \Pi^{\mu\nu}$, which might be 
helpful in order to further clarify the physics of damping for 
gluonic modes in the plasma. From it we easily obtain well known 
transversality properties as well as new constraints in particular 
concerning the imaginary part of $\displaystyle{ 
\raisebox{0.6ex}{\scriptsize{*}}} \Pi^{\mu\nu}$ 
related to damping. The expected equality $\gamma_l(m_g,0) = 
\gamma_t (m_g, 0)$ can also be derived. We report on these findings 
in the next two sections and further implications are discussed in 
the conclusions section. For convenience, the derivation of the ST 
identity is deferred to the Appendix. 
\setcounter{section}{1}
\section{Slavnov-Taylor identity and its consequences} 
Within the imaginary-time formalism of hot QCD, ST identities are 
derived from the BRST invariance of the Lagrangian in the same way 
as in vacuum theory. For a discussion of  dynamical quantities 
continuation to Minkowski space should be done in the usual way. 
We briefly summarize the procedure in the Appendix. 

Most important to the dynamics of gluonic fields is the identity 
involving the gluon polarization tensor $\Pi^{\mu\nu}$. In Coulomb 
gauge it reads: 
\begin{eqnarray} 
P_\nu \Pi^{\mu \nu} (P) & = & - \left[ \delta^\mu_{\; \, \nu} P^2 - 
P^\mu P_\nu + \Pi^\mu_{\; \, \nu} (P) \right] \Pi^\nu_g (P) \, , 
\label{ST10} 
\end{eqnarray} 
with $\Pi^\nu_g (P)$ related to the Coulomb-ghost self-energy part 
$\Pi_g (P)$ through $p^i \Pi^i_g (P) = \Pi_g (P)$.\footnote{Unless 
otherwise stated, throughout in this letter, Greek letters, $\mu$, 
$\nu$, ..., take $0,1, 2, 3$ while Ratin letters, $i$, $j$, ..., 
take $1, 2, 3$.} The derivation of the result (\ref{ST10}) is 
explained in the Appendix. Note that it is an exact identity, valid 
for arbitrary momenta $P$ and to all order in perturbation theory. 

>From Eq.~(\ref{ST10}), other identities are immediately obtained 
that will be useful for the discussion below. First contracting 
also the remaining Lorentz index we obtain 
\begin{eqnarray} 
P_\mu \Pi^{\mu \nu} (P) P_\nu & = & - P_\mu \Pi^\mu_{\; \, \nu} (P) 
\Pi_g^\nu (P) \nonumber \\ 
& = & [ P^2 g_{\mu \nu} - P_\mu P_\nu ] \, \Pi_g^\mu \Pi_g^\nu + 
\Pi^\mu_g \Pi_{\mu \nu} \Pi^\nu_g \, . 
\label{ST11} 
\end{eqnarray} 
Concerning the imaginary part of $\Pi^{\mu \nu} (P)$, 
since\footnote{This can readily be seen in real-time formalism 
\cite{lb}. The building blocks of perturbation theory, i.e., the 
propagators and vertices, and then also the self-energy part, are $2 
\times 2$ matrices in \lq\lq thermal space.'' $Im \Pi^\nu_g (P)$ 
here is proportional to the $(1, 2)$-component of $(\Pi^\nu_g 
(P))_{i j}$ $(i, j = 1, 2)$. Since, in Coulomb gauge, the ghost 
propagator-matrix $(\Delta_g (P))_{i j}$ $(i, j = 1, 2)$ is diagonal 
in thermal space, $(\Pi^\nu_g (P))_{i j}$ is also diagonal and 
$(\Pi^\nu_g (P))_{1 2} = 0$.} $Im \Pi_g^\nu (P) = 0$, we obtain 
\begin{eqnarray} 
P^\nu Im \Pi_{\mu \nu} (P) & = & - \Pi^\nu_g (P) Im \Pi_{\mu \nu} 
(P) \, . 
\label{ST10i} \\ 
P^\mu Im \Pi_{\mu \nu} (P) P^\nu & = & \Pi^\mu_g \Pi^\nu_g Im 
\Pi_{\mu \nu} \, . 
\label{ST11i} 
\end{eqnarray} 
Note again that, as Eq.~(\ref{ST10}), also Eqs.~(\ref{ST11}) - 
(\ref{ST11i}), are general, valid for arbitrary momenta and to all 
orders in perturbation theory. 

Eqs. (\ref{ST10})~-~(\ref{ST11i}) contain important informations. 
At leading and next to leading order several of them have been 
established through direct calculation. Consider the soft momentum 
region, $P^\mu = (\omega, {\bf p}) \sim gT$;  from Eqs.~(\ref{ST10}) 
and (\ref{ST11}) we readily rederive two well known results from 
power counting arguments: 
\begin{enumerate} 
\item At lowest nontrivial order of effective (HTL-resummed) 
perturbation theory \cite{htl,lb} $\Pi^{\mu \nu} (P)$ $(=$ $\delta 
\Pi^{\mu \nu} (P)) = O (g^2 T^2)$ so the left-hand side (LHS) of 
Eq.~(\ref{ST10}) is of $O (g^3 T^3)$. On the other hand $\Pi^\nu_g 
(P) = O (g^2 T)$ (no HTL-contribution in amplitudes with external 
ghosts \cite{htl,lb}) and the right-hand side (RHS) of 
Eq.~(\ref{ST10}) is of $O (g^4 T^3)$. Therefore, to the HTL 
accuracy, $P_\nu \Pi^{\mu \nu} (P) \simeq P_\nu \delta \Pi^{\mu \nu} 
(P) = 0$. 
\item Similarly, at leading order, the LHS of Eq.~(\ref{ST11}) is of 
$O (g^4 T^4)$. In the second line of Eq.~(\ref{ST11}), $\Pi^\mu_g 
(P) = O (g^2 T)$, $\Pi^{\mu \nu} (P) = \delta \Pi^{\mu \nu} (P) = O 
(g^2 T^2)$, $\Pi_g (P) = O (g^3 T^2)$. Therefore, RHS/LHS $= O 
(g^2)$, and 
\begin{equation} 
P_\nu \Pi^{\nu \mu} (P) P_\mu = 0 
\label{tra} 
\end{equation} 
holds to relative order $g^2$. A more precise statement is derived 
below, cf. Eq.~(\ref{ut}). 
\end{enumerate} 

We now proceed exploiting the results Eqs.~(\ref{ST10}) - 
(\ref{ST11i}) in more detail. We focus on the soft momentum region 
and the next to leading order contribution to $\Pi^{\mu \nu} (P)$ 
from which the leading damping behavior will be deduced. Following 
standard notations, we write $\Pi^{\mu \nu} (P)$ to this order 
$\delta \Pi^{\nu \mu} (P) + \displaystyle{ 
\raisebox{0.6ex}{\scriptsize{*}}} \Pi^{\nu \mu} (P)$ with $\delta 
\Pi$ the HTL-contribution and $\displaystyle{ 
\raisebox{0.6ex}{\scriptsize{*}}} \Pi$ the relative order $g$ 
correction.  The diagrams that yield leading contributions to the 
imaginary part of $\displaystyle{\raisebox{0.6ex}{\scriptsize{*}}} 
\Pi$ are shown in Fig.~1 below. 

In Eqs.~(\ref{ST10}) - (\ref{ST11i}) we need an explicit expression 
for the ghost contribution. It is sufficient to calculate $\Pi^\mu_g 
(P)$ to lowest nontrivial order, which we carry out in Appendix (cf. 
Eq.~(\ref{yareya})): 
\begin{eqnarray} 
\Pi^\mu_g (P) & \simeq & - \delta^{\mu i} \frac{g^2 N T}{16} 
\hat{p}^i \, , 
\label{yarer} 
\end{eqnarray} 
where $\hat{p}^i = p^i/p$. It is worth mentioning in passing that 
$\Pi^\mu_g (\omega, p = 0) = 0$. This is because, for ${\bf p} = 0$, 
the summand/integrand of Eq.~(\ref{6}) in Appendix, which is valid 
for arbitrary (Euclidean) four-momentum $P_E$, is odd in the 
summation/integration variables $K_E$. 

First we use Eq.~(\ref{yarer}) in Eq.~(\ref{ST10}): To next to 
leading order under consideration we can write  $\delta 
\Pi^\mu_{\; \, \nu}$ for $\Pi^\mu_{\; \, \nu}$ on the RHS of 
Eq.~(\ref{ST10}) and using $\delta \Pi^\mu_{\; \, \nu} P^\nu = 0$ 
\cite{htl,lb} as well as (\ref{tra}) we find 
\begin{eqnarray} 
P_\nu\displaystyle{ \raisebox{0.6ex}{\scriptsize{*}}} \Pi^{\nu \mu} 
(P) & = & \frac{g^2 N T}{16} \frac{\omega}{p} \displaystyle{ 
\raisebox{0.6ex}{\scriptsize{*}}} \! \Delta^{- 1}_l (P) \left( 1, \, 
\frac{\omega}{p} \hat{\bf p} \right)^\mu 
\label{main} 
\end{eqnarray} 
with the inverse plasmon propagator \cite{htl,lb}, 
\begin{eqnarray*} 
\displaystyle{ \raisebox{0.6ex}{\scriptsize{*}}} \! \Delta_l^{- 1} 
(P) & = & p^2 - \delta \Pi_{0 0} (P) = p^2 + 3 m_g^2 \left[ 1 - 
\frac{\omega}{2 p} \ln \left( \frac{\omega + p}{\omega - p} \right) 
\right] \, , 
\end{eqnarray*} 
where $m_g^2 = N (g T)^2 / 9$. Note that the close connection with 
the inverse propagator is of course apparent in Eq.~(\ref{ST10}). 
Also Eq.~(\ref{main}) is valid for any value of $\lambda$, the 
Coulomb gauge parameter, Eq.~(\ref{act}) (in Appendix). We have 
confirmed Eq.~(\ref{main}) through direct calculation of the LHS. 

Incidentally, from Eq.~(\ref{main}), $P_\mu \displaystyle{ 
\raisebox{0.6ex}{\scriptsize{*}}} \Pi^{\mu i} (P) e^{(r)}_i 
(\hat{\bf p}) = 0,~(r = 1, 2)$ for any three-vector $e_i^{(r)} ({\bf 
p})$ perpendicular to $\hat{\bf p}$ and arbitrary (soft) $P^\mu$. 

Using Eq.~(\ref{yarer}) in Eq.~(\ref{ST11}) we find 
\begin{equation} 
P_\mu \displaystyle{ \raisebox{0.6ex}{\scriptsize{*}}}\Pi^{\mu \nu} 
(P) P_\nu = - \left( \frac{g^2 N T \omega}{16} \right)^2 \left[1 - 
\frac{1}{\omega^2}\hat{p}^i \delta\Pi^{ij}(P) \hat{p}^j + O (g) 
\right] \, , 
\label{ut} 
\end{equation} 
which is of $O (g^6 T^4)$ when $P^\mu$ is soft, thus specifying 
somewhat the result under point (2) above. 

Finally, making progress towards the discussion of damping in the 
next subsection, we note that, from Eq.~(\ref{main}), the imaginary 
part of $ P_\mu \displaystyle{ \raisebox{0.6ex}{\scriptsize{*}}} 
\Pi^{\mu \nu} (P) $ is related to the Landau damping contribution in 
the HTL-propagator. We have: 
\begin{equation} 
Im \left[ P_\mu \displaystyle{ \raisebox{0.6ex}{\scriptsize{*}}} 
\Pi^{\mu \nu} (P) \right] = \frac{3 \pi}{3 2} g^2 N T m_g^2 \left( 
\frac{\omega}{p} \right)^2 \, \theta (p^2 - \omega^2) \, , 
\label{ima} 
\end{equation} 
and therefore for time-like momenta, 
\begin{equation} 
Im \left[ P_\mu \displaystyle{ \raisebox{0.6ex}{\scriptsize{*}}} 
\Pi^{\mu \nu} (P) \right] = 0 \;\;\;\;\; (P^2 > 0) \, . 
\label{ima1} 
\end{equation} 
\setcounter{section}{2} 
\section{Damping in a gluonic Medium} 
Soft gluon damping rates relate to the imaginary part of the gluon 
polarization tensor $\Pi^{\mu\nu}$, leading damping arises from the 
relative order g correction $\displaystyle{ 
\raisebox{0.6ex}{\scriptsize{*}}} \Pi^{\mu\nu}$. The relevant 
diagrams are shown in Fig.~1 where however the Coulomb-ghost loop 
does not contribute to $Im \Pi^{\mu \nu}$. HTL resummed propagators 
and vertices have to be used since the dominant contribution arises 
from configurations with soft loop momentum. For the following 
presentation we restrict to strict Coulomb gauge ($\lambda = 0$ in 
Eq.~(\ref{propgl}) in Appendix). 

For gluon damping only the transverse and longitudinal components of 
$\displaystyle{\raisebox{0.6ex}{\scriptsize{*}}} \Pi^{\mu\nu}$ 
\begin{eqnarray} 
\displaystyle{ \raisebox{0.6ex}{\scriptsize{*}}} \Pi_t (P) & = & 
\frac{1}{2} {\delta_\perp}^{i j} ({\hat{ \bf p}}) \displaystyle{ 
\raisebox{0.6ex}{\scriptsize{*}}} \Pi^{i j} (P) \, , \quad 
\delta_\perp^{i j} ({\hat {\bf p}}) = \delta^{i j} - {\hat p}^i 
{\hat p}^j  \, , \nonumber \\ 
\displaystyle{ \raisebox{0.6ex}{\scriptsize{*}}} \Pi_l (P) & = & - 
\displaystyle{ \raisebox{0.6ex}{\scriptsize{*}}} \Pi^{0 0} (P) \, . 
\label{longtrans} 
\end{eqnarray} 
will be relevant. In the vicinity of $\omega = \omega_{t / l} (p)$, 
the mass-shell for quasiparticles to HTL-accuracy, the respective 
inverse propagator component can be written: 
\begin{eqnarray*} 
& & G_{t / l}^{- 1}  \simeq   \mp Z_{t / l}^{- 1} (p) \, (\omega - 
\omega_{t / l}) + i Im \displaystyle{ 
\raisebox{0.6ex}{\scriptsize{*}}} \Pi_{t / l} (\omega = \omega_{t / 
l}, p) \, , \nonumber \\ 
& & Z_t^{- 1} (p) = \frac{3 m_g^2 \omega_t^2 - (\omega_t^2 - 
p^2)^2}{\omega_t (\omega_t^2 - p^2)} \, \stackrel{\tiny p\to 0}{\to} 
\, 2 m_g \, , \nonumber \\ 
&& Z_l^{- 1} (p) = 3 p^2 \left[ \frac{m_g^2}{\omega_l (\omega_l^2 - 
p^2)} - \frac{1}{3 \omega_l} \right] \, \stackrel{\tiny p\to 0}{\to} 
\, \frac{2 p^2}{m_g} \, . 
\end{eqnarray*} 
>From this, the damping rates for longitudinal and transverse gluonic 
modes are obtained as: 
\begin{eqnarray} 
\gamma_t (\omega_t (p), p) & = & -Z_t (p) \, Im \, \displaystyle{ 
\raisebox{0.6ex}{\scriptsize{*}}} \Pi_t (\omega_t (p), p) \quad 
\stackrel{\tiny p\to 0}{\to} \quad - \frac{1}{2m_g} \, Im \, 
\displaystyle{ \raisebox{0.6ex}{\scriptsize{*}}} \Pi_t (m_g, 0) \, , 
\nonumber \\  
\gamma_l (\omega_l (p), p) & = & + Z_l (p) \, Im \, \displaystyle{ 
\raisebox{0.6ex}{\scriptsize{*}}} \Pi_l (\omega_l (p), p) \quad 
\stackrel{\tiny p\to 0}{\to} \quad \frac{m_g}{2p^2} \, Im \, 
\displaystyle{ \raisebox{0.6ex}{\scriptsize{*}}} \Pi_l (m_g, 0)\, , 
\label{damp} 
\end{eqnarray}
with $m_g = \omega_{t/l}(p = 0)$ the plasma frequency. 

The direct calculation of $\gamma_{l/t}$, for general p, has not yet 
been accomplished, however a leading logarithmic singularity has 
been extracted for non-zero momentum in both $\gamma_t$ and 
$\gamma_l$ \cite{pis}. At zero momentum the calculation of 
$\gamma_t$ is well documented in the literature but note that 
calculating $\gamma_l$ requires to expand $ \displaystyle{ 
\raisebox{0.6ex}{\scriptsize{*}}} \Pi_l (P)$ to order $O(p^2)$ - 
a formidable and subtle task. A recent controversy \cite{al} 
suggests that approximations should be handled with care and 
observing general constraints. 

We are now in a position to show that our results so far leads to 
$\gamma_l (m_g, 0) = \gamma_t (m_g, 0)$ and thus a consistent 
calculation necessarily provides it. On the gluon mass-shell, $P^2 = 
\omega^2 - p^2 > 0$, we obtain from Eq.~(\ref{tra}) and 
Eq.~(\ref{ima1}) 
\begin{equation} 
Im \displaystyle{ \raisebox{0.6ex}{\scriptsize{*}}} {\Pi}^{0 0} (P) 
= - Im \displaystyle{ \raisebox{0.6ex}{\scriptsize{*}}} {\Pi}_l(P) = 
\left( \frac{p}{\omega} \right)^2 \hat{p}^i \hat{p}^j Im 
\displaystyle{ \raisebox{0.6ex}{\scriptsize{*}}} \Pi^{i j} (\omega, 
{\bf p}) \, . 
\label{pn0}
\end{equation}
Specializing to $p=0$ and using the fact {$\displaystyle{ 
\raisebox{0.6ex}{\scriptsize{*}}} \Pi^{i j} (\omega, {\bf p} = {\bf 
0}) \propto \delta^{i j}$} we find 
\begin{equation} 
\hat{p}^i \hat{p}^j Im \displaystyle{ 
\raisebox{0.6ex}{\scriptsize{*}}} {\Pi}^{i j} (\omega, {\bf 0}) = 
\frac{1}{3} \, Im \displaystyle{ \raisebox{0.6ex}{\scriptsize{*}}} 
{\Pi}^{i i} (\omega, {\bf 0}) = Im \displaystyle{ 
\raisebox{0.6ex}{\scriptsize{*}}} {\Pi}_t (\omega, {\bf 0}) \, . 
\label{p0} 
\end{equation} 
Finally, using  $\omega_{l / t} (p = 0) = m_g$ and Eq.~(\ref{pn0}) 
with (\ref{p0}) in Eq.~(\ref{damp}), we derive that 
\begin{equation} 
\gamma_l (m_g, 0) = \gamma_t (m_g, 0) 
\label{eq} 
\end{equation} 
holds, as is expected on physical grounds. 

Finally it is interesting to observe what consequences arise if we 
assume, on physical grounds, that $\gamma_l(m_g, 0) = \gamma_g(m_g, 
0)$ to all orders in perturbation theory. From 
Eqs.~(\ref{longtrans}), (\ref{damp}) this assumption leads to  
\[ 
\lim_{p\to0} \left(\frac{\omega}{p}\right)^2 Im \Pi^{0 0} = 
\hat{p}^i \hat{p}^j  Im \Pi^{i j} (\omega, \bf{0}), 
\] 
and using this in Eq.~(\ref{ST11i}) one finds 
\[ 
\lim_{p \to 0} \frac{1}{p^2} \Pi^\mu_g [\Pi^\nu_g + 2 \delta^{\nu i} 
\, p^i] \, Im \Pi_{\mu \nu} = 0 \, . 
\] 
\section{Conclusions and remarks} 
The Slavnov-Taylor identity for gluon polarization tensor we report 
on in this letter, Eq.~(\ref{ST10}), allows to obtain various 
constrains. Applying it to the next to the leading order in 
HTL-resummed perturbation theory, the identity Eq.~(\ref{main}) 
results, which is relevant to leading-order damping rate of gluonic 
modes. As has been discussed, for zero three momentum the expected 
equality $\gamma_t (m_g,0) = \gamma_l(m_g,0)$ can be derived with 
the help of the identity Eq.~(\ref{main}). One of the advantages of 
the above procedure is that the explicit expansion of 
$\displaystyle{ \raisebox{0.6ex}{\scriptsize{*}}} \Pi^{0 0} 
(\omega_l, p)$ around $p \sim 0$, which has been found to be 
troublesome, can be avoided in this derivation. Moreover, concerning 
the direct calculation of $\gamma_l$, the present work indicates 
that dealing with singularities in intermediate steps of the 
calculation and necessary changes in the integration variables 
should be made in a way consistent with the identity 
Eq.~(\ref{main}). 
\setcounter{section}{1}
\section*{Appendix Derivation of (1) and (6)} 
We start in the  imaginary-time formalism continuing to real 
energies at the final stage. The QCD action in Coulomb gauge reads 
\begin{eqnarray} 
S & = & \int_0^{1 / T} d x_0 \int d^{\, 3} x {\cal L} (x) \;\; 
\left( \equiv \int_T d^{\, 4} x \, {\cal L} (x) \right) \, , 
\nonumber \\ 
{\cal L} (x) & = & - \frac{1}{4} F^{\mu \nu}_a (x) F^{\mu \nu}_a (x) 
- \frac{1}{2 \lambda} (\nabla \cdot {\bf A}_a (x))^2 + \delta^{\mu 
i} \bar{\eta}_a (x) \partial^i D_{x, a b}^\mu (A) \eta_b (x) \, , 
\nonumber \\ 
F^{\mu \nu}_a & = & \partial^\mu A_a^\nu - \partial^\nu A_a^\mu + g 
C_{a b c} A_b^\mu A_c^\nu \, . 
\label{act} 
\end{eqnarray} 
Here $D_{x, a b}^\mu (A) \equiv \delta_{a b} \partial / \partial 
x^\mu - g C_{a b c} A_c^\mu (x)$ and $\bar{\eta}_a$ and $\eta_a$ are 
the Coulomb-ghost fields. For notations related to the color space, 
we follow \cite{IZ}. 

The bare gluon propagator $\tilde{\Delta}^{\mu \nu} (P_E)$ reads 
\begin{equation} 
\tilde{\Delta}^{\mu \nu} (P_E) = \delta^{\mu i} \delta^{\nu j} 
\frac{\delta^{i j} - \hat{p}^i \hat{p}^j}{P_E^2} + \delta^{\mu 0} 
\delta^{\nu 0} \frac{1}{p^2} + \lambda \frac{P_E^\mu P_E^\nu}{p^4} 
\, , 
\label{propgl} 
\end{equation} 
where $P_E^\mu = (p_0, {\bf p})$ with $p_0 = 2 \pi T n$ $(n = 0 \pm 
1, \pm 2, ...)$, and the bare ghost propagator reads 
$\tilde{\Delta}_g (p) = 1 / p^2$. The form of gluon-ghost vertex can 
be read off from ${\cal L} \ni g C_{a b c} (\partial^j \bar{\eta}_a) 
\eta_b A^j_c$, 
\begin{equation} 
{\cal V}_g = - i g C_{a b c} \, p^j \, , 
\label{ver} 
\end{equation} 
where ${\bf p}$ is the outgoing momentum carried by $\bar{\eta}_a$. 

The generating functional reads 
\begin{eqnarray} 
Z [J, \bar{\xi}, \xi] & = & \int {\cal D} A^\mu_a {\cal D} \eta_a 
{\cal D} \bar{\eta}_a \exp \left[ S + \int_T d^{\, 4} x \left\{ 
J^\mu_a (x) A^\mu_a (x) \right. \right. \nonumber \\ 
&& \left. \left. + \bar{\xi}_a (x) \eta_a (x) + \bar{\eta}_a (x) 
\xi_a (x) \right\} \right]  \, , 
\label{gene} 
\end{eqnarray} 
where, the functional integral is to be performed with periodic 
boundary conditions for all fields (including ghosts \cite{kugo}), 
$A^\mu_a (x_0 = 0, {\bf x}) = A^\mu_a (x_0 = 1 / T, {\bf x})$, etc. 

The action $S$ (or ${\cal L}$), Eq.~(\ref{act}), is invariant under 
the BRST transformation \cite{IZ}: 
\[ 
\delta A_a^\mu = D_{a b}^\mu (A) \eta_b \, \delta \zeta \, , \quad 
\delta \bar{\eta}_a = - \frac{1}{\lambda} \nabla \cdot {\bf A}_a \, 
\delta \zeta \, , \quad \delta \eta_a = \frac{g}{2} C_{a b c} \eta_b 
\eta_c \, \delta \zeta \, , 
\] 
where $\delta \zeta$ is a Grassmann-number parameter. Using this 
fact in (\ref{gene}), we obtain, 
\begin{eqnarray} 
&& \int_T d^{\, 4} z {\cal B} (z) Z [J, \bar{\xi}, \xi] = 0 \, , 
\nonumber \\ 
&& {\cal B} (z) = J^\mu_a (z) D_{z, a b}^\mu \left( 
\frac{\delta}{\delta J} \right) \frac{\delta}{\bar{\xi}_b (z)} 
+ \frac{1}{\lambda} \xi_a (z) \frac{\partial}{\partial z^i} 
\frac{\delta}{\delta J_a^i (z)} \nonumber \\ 
&& \mbox{\hspace*{7ex}} + \frac{g}{2} C_{a b c} \bar{\xi}_a (z) 
\frac{\delta}{\delta \bar{\xi}_b (z)} \frac{\delta}{\delta 
\bar{\xi}_c (z)} \, . 
\label{tane} 
\end{eqnarray} 

Computing 
\[ 
\frac{\delta}{\delta J_a^\mu (x)} \frac{\delta}{\delta \xi_b (y)} 
\int_T d^{\, 4} z {\cal B} (z) \ln Z [J, \bar{\xi}, \xi] \, 
\rule[-3mm]{.14mm}{8.5mm} \raisebox{-2.85mm}{\scriptsize{$\; J = 
\bar{\xi} = \xi = 0$}} \, , 
\] 
by using (\ref{tane}), we obtain 
\begin{equation} 
\delta_{a b} \partial^\mu_x \tilde{G}_g (x - y) - g C_{a c d} 
\langle A^\mu_d (x) \eta_c (x) \bar{\eta}_b (y) \rangle + \delta_{a 
b} \frac{1}{\lambda} \frac{\partial}{\partial y^i} \tilde{G}^{\mu i} 
(x - y) = 0 \, . 
\label{s2} 
\end{equation} 
Note that the third term of Eq.~(\ref{tane}) does not contribute to 
(\ref{s2}). The ghost propagator $\tilde{G}_g$ and the gluon 
propagator $\tilde{G}^{\mu \nu}$ in Eq.~(\ref{s2}) are defined, 
respectively, through 
\begin{eqnarray*} 
\delta_{a b} \tilde{G}_g (x - y) & = & \langle \eta_a (x) 
\bar{\eta}_b (y) \rangle \equiv \frac{\delta \ln Z}{\delta \xi_b (y) 
\delta \bar{\xi}_a (x)} \,  \rule[-3mm]{.14mm}{8.5mm} 
\raisebox{-2.85mm}{\scriptsize{$\; J = \bar{\xi} = \xi = 0$}} \, , 
\nonumber \\ 
\delta_{a b} \tilde{G}^{\mu \nu} (x - y) & = & \langle A^\mu_a (x) 
A^\nu_b (y) \rangle \equiv \frac{\delta \ln Z}{\delta J^\mu_a (x) 
\delta J^\nu_b (y)}\, \rule[-3mm]{.14mm}{8.5mm} 
\raisebox{-2.85mm}{\scriptsize{$\; J = \bar{\xi} = \xi = 0$}} \, . 
\end{eqnarray*} 
$\langle A^\mu_d (x) \eta_c (x) \bar{\eta}_b (y) \rangle$ is the 
gluon-ghost three-point function. 

Now, we call for the Schwinger-Dyson equations: 
\begin{eqnarray} 
\tilde{G}^{\mu \nu} (x - y) & = & \tilde{\Delta}^{\mu \nu} (x - y) 
\nonumber \\ 
& & -  \int_T d^{\, 4} u d^{\, 4} v \, \tilde{G}^{\mu \rho} (x - u) 
\tilde{\Pi}^{\rho \sigma} (u - v) \tilde{\Delta}^{\sigma \nu} (v - 
y) \, , 
\label{SD1} \\ 
\delta_{a b} \tilde{G}_g (x - y) & = & \delta_{a b} \tilde{\Delta}_g 
(x - y) \nonumber \\ 
& & + g C_{a d e} \int_T d^{\, 4} z \frac{\partial \tilde{\Delta}_g 
(x - z)}{\partial z^i} \langle A_e^i (z) \eta_d (z) \bar{\eta}_b (y) 
\rangle \, , 
\label{SD2} 
\end{eqnarray} 
where $\tilde{\Pi}^{\rho \sigma}$ is the gluon polarization tensor. 
It can readily be shown that, in the limit $g \to 0$, (\ref{s2}) 
holds, as it should be. Taking this fact into account, we substitute 
(\ref{SD1}) and (\ref{SD2}) into (\ref{s2}) and use the form 
(\ref{propgl}) for $\tilde{\Delta}^{\sigma i}$ to obtain 
\begin{eqnarray} 
&& \delta_{a b} \int_T d^{\, 4} z \tilde{\Pi}^{\nu \mu} (x - z) 
\partial^\mu_y \hat{\Delta}_g (z - y) \nonumber \\ 
&& \mbox{\hspace*{4ex}} = - g C_{a c d} \int_T d^{\, 4} \xi \int_T 
d^{\, 4} z (\tilde{G}^{- 1})^{\nu \mu} (x - \xi) \nonumber \\ 
&& \mbox{\hspace*{7ex}} \times \left[ \delta^{\rho i} 
\partial_\xi^\mu \partial_\xi^i \tilde{\Delta}_g (\xi - z) + 
\delta^{\, 4} (\xi - z) \delta^{\mu \rho} \right] \langle A_d^\rho 
(z) \eta_c (z) \bar{\eta}_b (y) \rangle \, . 
\label{s3} 
\end{eqnarray} 
The inverse propagator $(\hat{G}^{- 1})^{\nu \mu}$ is related to 
$\tilde{\Pi}^{\nu \mu}$ through $(\tilde{G}^{- 1})^{\nu \mu} = 
(\tilde{\Delta}^{- 1})^{\nu \mu} + \tilde{\Pi}^{\nu \mu}$. The 
Fourier transform of $\langle A_d^\rho (z) \eta_c (z) \bar{\eta}_b 
(y) \rangle$ may be written as 
\begin{equation} 
g C_{a c d} \left[ \langle A_d^\rho (z) \eta_c (z) \bar{\eta}_b (y) 
\rangle \right]_{\mbox{\scriptsize{F.T.}}} = - i \delta_{a b} 
\tilde{\Pi}^\rho_g (P_E) \tilde{G}_g (P_E) \, , 
\label{sd1} 
\end{equation} 
where $( \tilde{\Pi}^\rho_g (P_E) )_{a b}$ is related to a ghost 
self-energy part $\tilde{\Pi}_g$ through 
\begin{equation} 
p^i \tilde{\Pi}^i_g (P_E) = \tilde{\Pi}_g (P_E) \, . 
\label{gs} 
\end{equation} 

Putting altogether in (\ref{s3}) and transforming to Fourier space, 
we finally obtain 
\begin{eqnarray} 
\tilde{\Pi}^{\nu \mu} (P_E) P_E^\mu & = & - p^2 \tilde{G}_g (P_E) 
\left[ \left\{ \delta^{\nu \mu} P_E^2 - P_E^\nu P_E^\mu + 
\tilde{\Pi}^{\nu \mu} (P_E) \right\} \tilde{\Pi}^\mu_g (P_E) 
\right. \nonumber \\ 
&& \left. - \frac{P_E^\mu}{p^2} \tilde{\Pi}^{\mu \nu} (P_E) 
\tilde{\Pi}_g (P_E) \right] \, , 
\label{s41} 
\end{eqnarray} 
where use has been made of (\ref{gs}). Noting that $\tilde{G}_g 
(P_E) = [p^2 + \tilde{\Pi}_g (P_E)]^{- 1}$ and using 
$\tilde{\Pi}^{\mu \nu} = \tilde{\Pi}^{\nu \mu}$, Eq.~(\ref{s41}) 
may be \lq solved' as 
\begin{eqnarray} 
P_E^\nu \tilde{\Pi}^{\mu \nu} (P_E) & = & - \left[ \delta^{\mu \nu} 
P_E^2 - P_E^\mu P_E^\nu + \tilde{\Pi}^{\mu \nu} (P_E) \right] 
\tilde{\Pi}^\nu_g (P_E) \, . 
\label{s4} 
\end{eqnarray} 
>From this we obtain Eq.~(\ref{ST10}) in the main text after 
continuation to real energies $i p_0 \to \omega + i 0^+$ according 
to 
\begin{eqnarray} 
&& P^\mu = (\omega, {\bf p}) \, , \;\;\;\; P_E^2 \to - P^2 = p^2 - 
\omega^2 \, , \nonumber \\ 
&& \tilde{\Pi}^{\mu \nu} (i p_0, {\bf p}) \to (- i)^{\delta_{\mu 0} 
+ \delta_{\nu 0}} \Pi^{\mu \nu} (\omega, {\bf p}) \, , \;\;\;\;\; 
\tilde{\Pi}_g^{\mu} (i p_0, {\bf p}) \to (- i)^{\delta_{\mu 0}} 
\Pi_g^{\mu} (\omega, {\bf p}) \, . \nonumber \\ 
\label{Apfinal} 
\end{eqnarray} 

Finally we compute $\Pi_g^{\mu} (P)$ with soft $P$ to lowest 
non-trivial order --- one-loop contribution --- as 
required for the discussion in the main text. Using 
Eqs.~(\ref{propgl}) and (\ref{ver}), we have 
\begin{eqnarray} 
\tilde{\Pi}^\mu_g (P_E) & = & - g^2 N \, T 
\sum_{k_0}\int\frac{d^3 k}{(2\pi)^3} \, \frac{(p^i - k^i)}{({\bf p} 
- {\bf k})^2} \tilde{\Delta}^{i \mu} (K_E) \nonumber \\ 
& = & - g^2 N \, T \sum_{k_0} \int\frac{d^3 k}{(2\pi)^3} \, 
\frac{p^i - k^i}{({\bf p} - {\bf k})^2} \left[ \delta^{\mu 
j} \delta_\perp^{i j} (\hat{\bf k}) \tilde{\Delta} (K_E) + \lambda 
\frac{k^i K_E^\mu}{k^4} \right] \, , 
\label{6} 
\end{eqnarray} 
where $p^i = (0, {\bf p})$, $K_E^\mu = (k_0, {\bf k})$ and 
${\delta_\perp}^{i j} ({\hat{ \bf k}})$ is as in (\ref{longtrans}). 
In Eq.~(\ref{6}), $\tilde{\Delta} (K_E) = 1 / K_E^2$ for hard $p$, 
while, for soft $p$, the HTL-resummed $\displaystyle{ 
\raisebox{0.6ex}{\scriptsize{*}}} \! \tilde{\Delta} (K_E)$ is 
substituted for $\tilde{\Delta} (K_E)$. After carrying out the 
renormalization, one can easily see that the leading contribution 
comes from the first term (in the square brackets) with soft $k$ 
region: 
\begin{eqnarray} 
\tilde{\Pi}^\mu_g (P_E)& \simeq & - g^2 N \, T \sum_{k_0} 
\int_{\mbox{\scriptsize{soft $k$}}} \frac{d^3 k}{(2\pi)^3} \, 
\frac{(p^i - k^i)}{({\bf p} - {\bf k})^2} \delta^{\mu j} 
\delta_\perp^{i j} (\hat{\bf k}) \nonumber \\ 
& & \qquad \times \int_0^{1 / T} d \tau e^{i k_0 \tau} \int d \zeta 
\, \rho_t (\zeta, k) [1 + n (\zeta)]  e^{- \zeta \tau} \, , 
\label{Apghost}
\end{eqnarray}
where $\rho_t (\zeta, k)$ is the spectral function of the 
HTL-resummed transverse gluon propagator and $n (\zeta) = 1 / 
(e^{\zeta / T} - 1)$. On summing over $k_0$ and using $n (\zeta) 
\simeq T / \zeta$ and $\int d \zeta \rho_t (\zeta, k) / \zeta = 1 / 
k^2$, we obtain 
\begin{eqnarray} 
\tilde{\Pi}^\mu_g (P_E) & \simeq & - g^2 N T \delta^{\mu i} 
\int_{\mbox{\scriptsize{soft $k$}}} \frac{d^{\, 3} k}{(2 \pi)^3} 
\frac{p^i - ({\bf p} \cdot \hat{\bf k}) \hat{k}^i}{k^2 ({\bf p} - 
{\bf k})^2} \nonumber \\ 
& \simeq & - \delta^{\mu i} \frac{g^2 N T}{16} \hat{p}^i \, , 
\label{yareya} 
\end{eqnarray} 
which is independent of $\lambda$. Noting that $\tilde{\Pi}^\mu_g$ 
in (\ref{yareya}) has only spatial components and is independent of 
$p_0$, we have $\Pi^\mu_g (P) = \tilde{\Pi}^\mu_g (P_E)$ (cf. 
Eq.~(\ref{Apfinal})). 
\section*{Acknowledgements} 
M. D. likes to thank the German Academic Exchange Office (DAAD) for 
financial and general support of his stay in Osaka. Thanks also to 
the Faculty of Science, Osaka City University, for kind hospitality. 
 
\newpage

\begin{figure}
\centerline{\epsfig{file=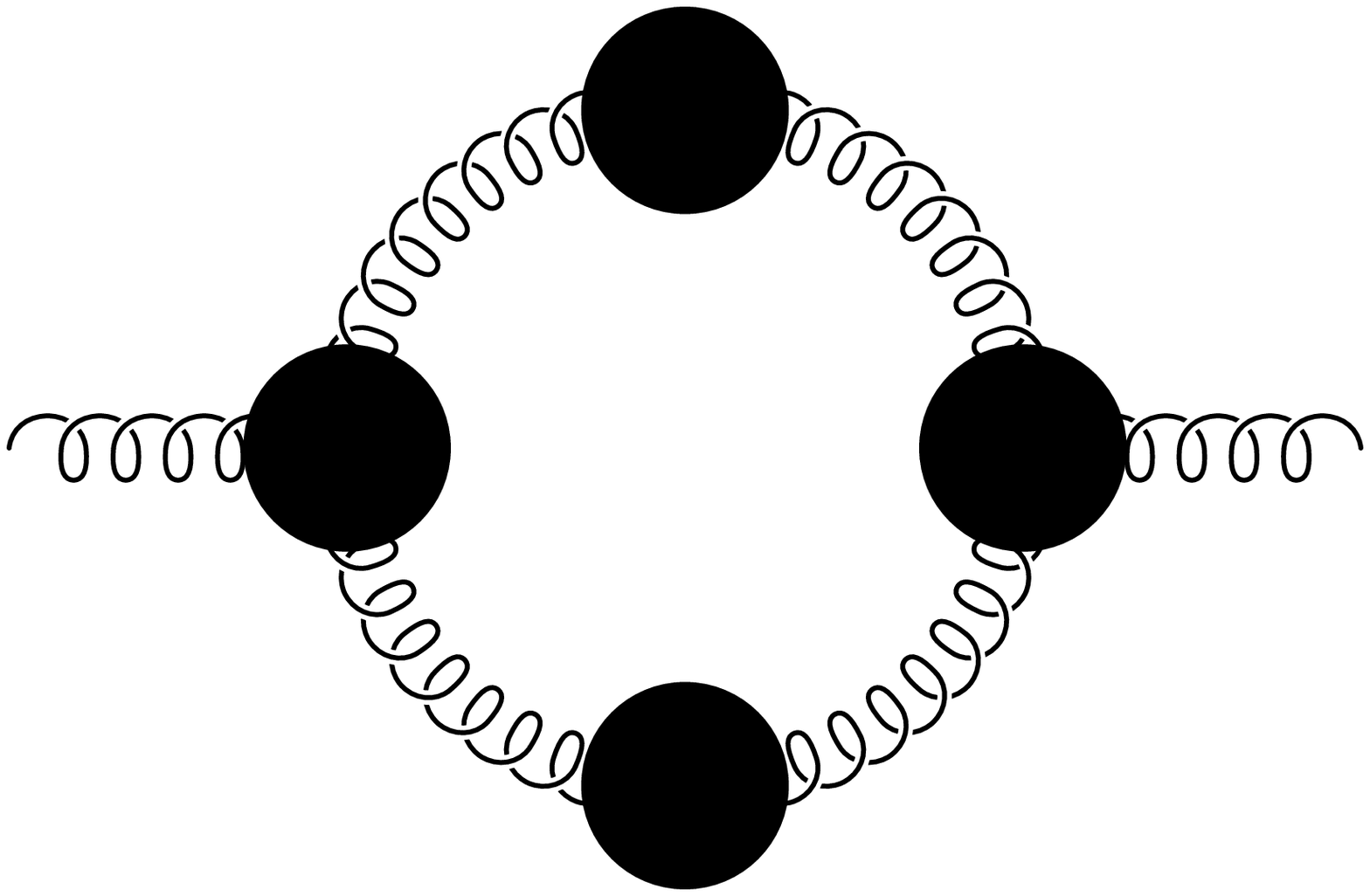,width=5cm}\qquad\qquad
\epsfig{file=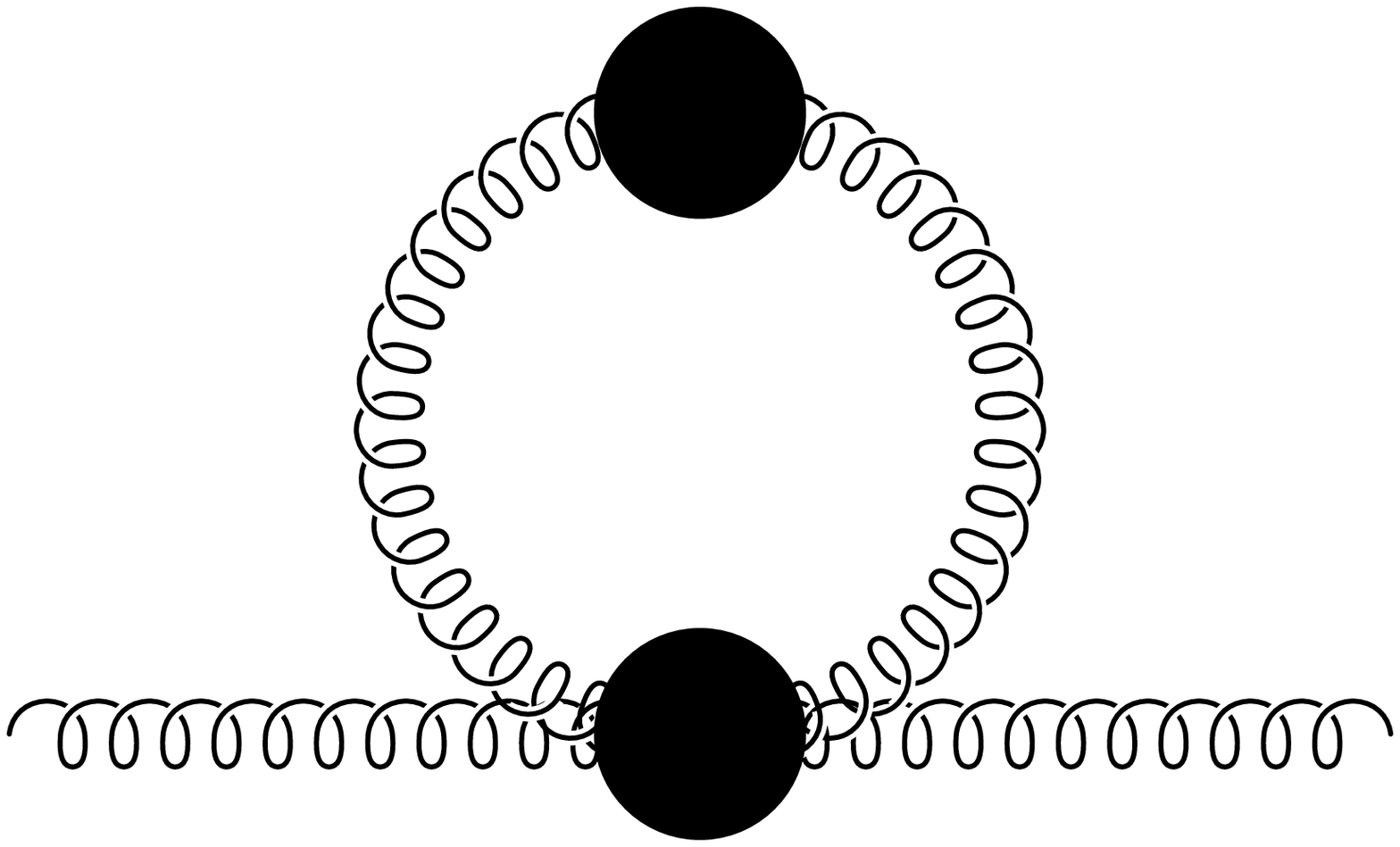, width=5cm}}
\caption{\label{fig}Next to leading order contributions to the
imaginary part of the
gluon polarization tensor. All loop-momenta are soft and HTL-resummed
propagators and vertices have to be used.}     
\end{figure}

\end{document}